\documentclass[useAMS,usenatbib,usegraphicx]{mn2e}
\usepackage{wasysym}
\title [New prospects for observing and cataloguing exoplanets in well detached binaries]
{New prospects for observing and cataloguing exoplanets in well detached binaries}
\author[R. Schwarz, B. Funk, R. Zechner and \'A. Bazs\'o]
{R. Schwarz\thanks{E-mail:schwarz@astro.univie.ac.at}, B. Funk, R. Zechner and \'A. Bazs\'o\\
Institute for Astronomy, University of Vienna, A-1180 Vienna, 
T\"urkenschanzstrasse 17, Austria\\ 
}

\begin{document}

\date{Accepted 1988 December 15. Received 1988 December 14; in original 
form 1988 October 11}

\pagerange{\pageref{firstpage}--\pageref{lastpage}} \pubyear{2002}
\maketitle
\label{firstpage}

\begin{abstract}
This paper is devoted to study the circumstances favourable to detect circumstellar
and circumbinary planets in well detached binary-star-systems using eclipse timing variations 
(ETVs). 
We investigated the dynamics of well detached binary star systems  with a star separation 
from 0.5 to 3~AU, to determine the probability of the detection of such variations with ground 
based telescopes and space telescopes (like former missions CoRoT and Kepler 
and future space missions Plato, Tess and Cheops). For the chosen star separations
both dynamical configurations (circumstellar and circumbinary) may be observable.
We performed numerical simulations by using the full three-body problem 
as dynamical model. The dynamical stability and the ETVs are investigated by computing 
ETV maps for different masses of the secondary star and the exoplanet (Earth, Neptune 
and Jupiter size). In addition we changed the planet's and binary's eccentricities. 
We conclude that many amplitudes of ETVs are large enough to detect exoplanets in 
binary star systems. 
As an application, we prepared statistics of the catalogue of exoplanets in binary star systems 
which we introduce in this article and compared the statistics with our parameter-space 
which we used for our calculations. In addition to these statistics of the catalogue 
we enlarged them by the investigation of well detached binary star systems from 
several catalogues and discussed the possibility of further candidates.
\end{abstract}


\section{Introduction}
\label{intro}

The first extra solar planet was discovered in the early 1990s by 
\citet{w}. Today the statistics of the observations show that the 
architecture of our solar system seems to be unique compared with exoplanetary 
systems. At the moment we know about 2000 exoplanets in more than 1200 
planetary systems, among them more than 100 exoplanets are in binary-star 
systems and two dozen are in multiple-star systems. The data of all planets are 
collected in the Exoplanet-catalogue maintained by J. 
Schneider\footnote{http://exoplanet.eu}; 
whereas the binary and multiple-star systems can be found separately in the 
catalogue of exoplanets in binary star systems
\footnote{http://www.univie.ac.at/adg/schwarz/multiple.html} 
maintained by R. Schwarz, which we will also introduce in this paper.\\

Approximately 70 percent of the main- and pre-main-sequence stars are members of 
binary or multiple star systems: 67 \% for G-M star, e.g. \citet{mayor}; and approximately 
70 \% for O-B stars (e.g. \citet{fabricius}, \citet{sana2012}). Statistics of solar-type 
dwarfs were studied by \citet{toko14} with a distance-limited
sample of 4847 targets. A field population was found of about 54\% for single stars,
33\% binary stars, 8\% triple systems, 4\% for quadrupole systems, 1\% for systems $N>4$. 
Observational evidence indicates that many of these systems contain potentially 
planet-forming circumstellar or circumbinary discs, implying that planet formation may 
be a common phenomenon in and around binary stars (e.g. \citet{mathieu1994}, 
\citet{akeson1998}, \citet{rodriguez1998}, \citet{trilling2007}). 
This fact led many research groups to examine the planetary formation and
evolution and dynamical stability in binary star systems, either in general or for selected 
systems \citep{andra15, dvorak03, hagh06, hagh10, holman97, kley08, mus05, paar08, pl02, 
pl03, ragh06, sal09, tak08, the10}. Despite many theoretical studies on the planetary 
formation in double star systems, the formation processes are not entirely understood 
\citep{kley, brom, jang, markus14}.\\

From the dynamical point of view the binary star systems as well as multiple star
systems are particularly interesting.
According to the work of \citet{rabl} one can distinguish three types of 
planetary orbits in a binary star system:
\begin{enumerate}
\item S-Type or circumstellar motion, where the planet orbits one of the two stars;
\item P-Type or circumbinary motion, where the planet orbits the entire binary;
\item T-Type: a planet may orbit close to one of the two equilibrium points 
$L_4$ and $L_5$; we call them Trojan planets. The dynamical study of \citet{sch09} could show
 with a few real binary systems that the T-Type configuration 
is not only of theoretical interest and \citet{sch15} could show that T-type orbits can be 
detected with ETV signals. 
\end{enumerate}

\begin{figure}
\centerline
{\includegraphics[width=9.5cm,angle=0]{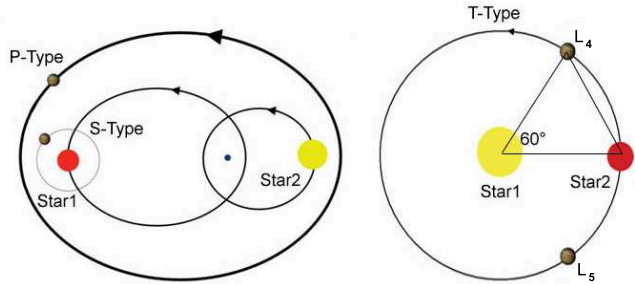}}
\caption{Scheme of the 3 different dynamical possibilities of planets in double stars. A colour 
version of this figure is available in the online version.}
\label{fig1}
\end{figure}

The graphic representation of the different dynamical scenarios is given in Fig.~\ref{fig1}.
The first planet in P-Type motion, was detected in 2009 \citep[HW Vir (AB) c, ][]{lee2009}
\footnote{Newer investigations of \citep{funk} and \citep{horner12} found that this system is 
controversial.}. 
Since that time planets in well detached binary systems become more and more attractive, especially
tight coplanar circumbinary planets around short-period binaries \citep{ham}.
Further P-Type planets were discovered in the following years, where especially the 
space-mission Kepler was very successful. Among them are also multiplanetary circumbinary 
systems, like HW Virginis or Kepler 47 \citep{orosz2012}.\\

From the observational point of view well detached binary star systems with separations smaller 
than 3 AU are more interesting than wide binary systems because the observation time for the 
latter ones is much longer.
Furthermore, well detached binaries offer reasonable signal-to-noise ratio (S/N) values for 
photometry and radial velocity (RV) amplitudes \citep{guedes,beauge,malbet,eggl13}.

A first study of test particles in circumbinary orbits was presented by
\citet{dvorak1986}, \citet{dvorak1989} and \citet{holman1999}. \citet{schwarz2011} 
studied the dynamics of binary star systems with a circumbinary planet, and calculated 
its eclipse timing variations (ETVs) for different values of the mass ratio and orbital 
elements of the binary and the perturbing body.\\

\begin{figure}
\centerline
{\includegraphics[width=5.8cm,angle=270]{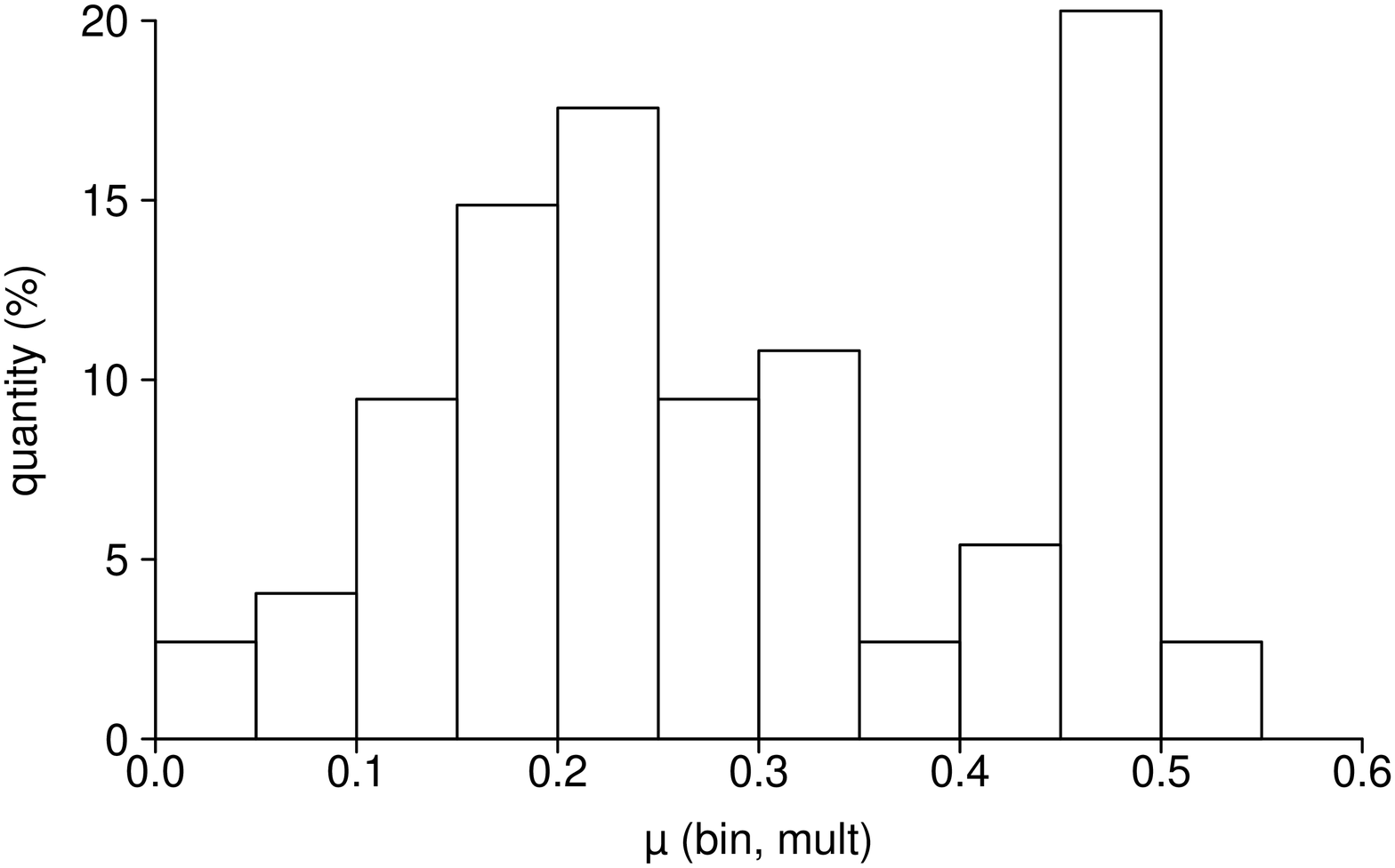}}
{\includegraphics[width=8.7cm,angle=0]{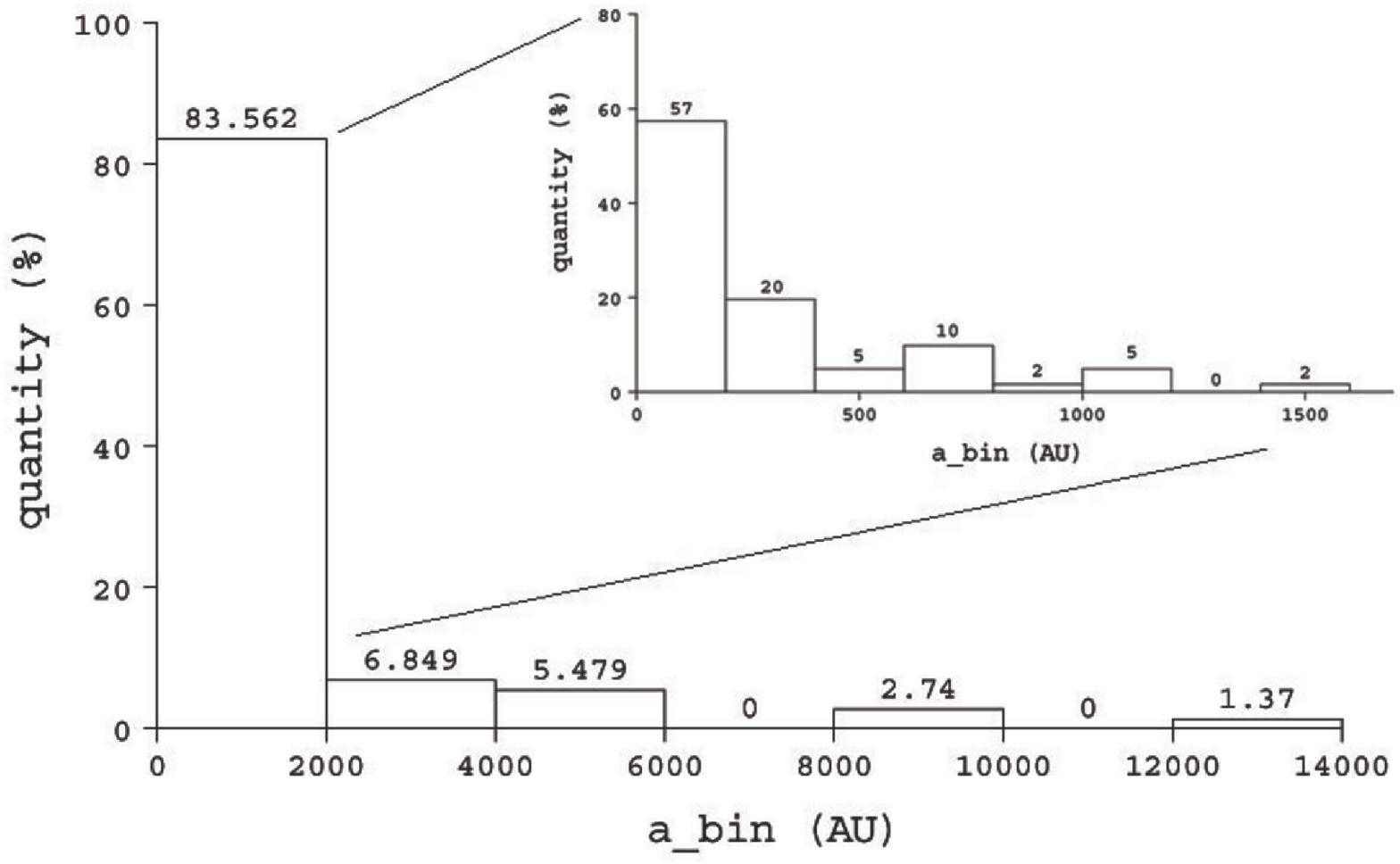}}
\caption{The histogram presents the mass ratio $\mu_{bin}$ (upper graph) and the separation 
($a_{bin}$, shown in the lower graph) of all binary star systems with exoplanets (including the 
binaries in multiple star systems, see Figs.~\ref{triple},~\ref{quad}), taken from the binary 
catalogue of exoplanets (http://www.univie.ac.at/adg/schwarz/multiple.html). To see the well 
detached binary systems we zoom into the histogram of $a_{bin}$ in the inset.}
\label{fig2}
\end{figure}

Most observations of planets in binaries are focused on $\mu\approx 0.5$ 
(stars have similar masses) and are restricted to Sun-like stars. 
In Fig.~\ref{fig2} we show a distribution of the mass ratios of all detected exoplanets 
in binaries and we found that the most common mass ratios $\mu ={\frac {m_2} {(m_1+m_2)}}$ 
are $\mu=0.25$ and 0.5. 
Therefore we use different mass ratios for our simulations for P- and S-Type systems.

This paper is divided into three parts: the first part is devoted to the possible detection 
of exoplanets in well detached binary star systems in P- and S-Type motion by the help of eclipse 
timing variations (ETV).
In the second part we prepare statistics for well detached binary star systems from several 
catalogues and discussed the possibility of further candidates.
The actual statistics of planets in binaries and multiple star systems are taken from the 
catalogue of exoplanets in binary star systems which we introduce in the chapter 6.

\section{Numerical setup}
\subsection{Models}
\label{models}
The photometric detection of extrasolar planets is of particular interest
for the discoveries in eclipsing binaries. We investigated well detached binary star systems, 
where the initial separation of the stars is 0.5 to 3~AU.
From the dynamical point of view these initial separations are very interesting, because planets in
S- and P-Type orbits are possible and they are supported by the first Kepler discoveries of a long 
($\approx 400^d$) eclipsing binary with a pulsating red giant component~\citep{hekker}

We studied the planar full three-body problem (3BP) with numerical
integrators. In this problem two finite bodies, the primaries ($m_1=primary$ $star$, 
$m_2=secondary$ $star$) revolve about their common center of mass, starting 
with three different eccentricities ($e_2=0, 0.2, 0.4$). 
A third body $m_{3}=planet$ moves around $m_1$ for S-Type or around 
both stars for P-Type motion in the same plane as $m_2$.
\par
We have regarded all the celestial bodies involved as point masses and
integrated the equations of motion for a time up to $T_c=10^3$~yrs for 
the ETV-maps shown in section~\ref{S} and~\ref{P} and $T_c=10^6$~yrs for the stability limits.
For our simulations we used a Gauss Radau integrator with an adaptive
step size \citep[see][and references therein]{eggl} for the ETVs and for the stability the 
Lie-method with an automatic step size control to solve the equations of motion.
In the S-Type configuration we also investigated close-in planets, therefore we considered 
the general relativity in our calculations, \citet{beutler}.

We considered well detached eclipsing binaries, with distances between the
stars of 0.5, 1 and 3~AU. To ensure that the effect of the variations
of the mass ratio $\mu$ of the binary is included in 
our study, we considered the following three models:
\begin{itemize}
  \item {\bf model 1:} $m_1$ = $m_2$ = 1$M_{\odot}$, corresponds to $\mu=0.5$
  \item {\bf model 2:} $m_1$ = 1$M_{\odot}$ and $m_2$ = 0.5$M_{\odot}$, corresponds to $\mu=0.33$
  \item {\bf model 3:} $m_1$ = 1$M_{\odot}$ and $m_2$ = 0.1$M_{\odot}$, corresponds to $\mu=0.09$
\end{itemize}

As shown in Fig.~\ref{fig2} the mass ratio of our models is quite common, 
when we look at the histogram of the detected exoplanets in binaries and multiple star systems.
We changed the other two models ($\mu=0.33$ and $\mu=0.09$) because the statistics of other binary
catalogues (section\ref{statistics}) are equally-distributed.
To get a good estimation about occurring perturbations on the secondary star
(to measure ETVs) we used planets with different masses $m_3$: 
Earth\footnote{$M_{\oplus}$ which corresponds to $3 \cdot 10^{-6}$ $M_{\odot}$}, 
Neptune\footnote{$M_{\neptune}$ which corresponds to $5 \cdot 10^{-5}$ $M_{\odot}$},
and Jupiter\footnote{$M_{\jupiter}$ which corresponds to $1 \cdot 10^{-3}$ $M_{\odot}$}.

\subsection{Methods} 
For the analysis of the orbit we used the method of the maximum 
eccentricity $e_{max}$. In former studies we found a good agreement with 
chaos indicators like the Lyapunov characteristic indicator (LCI)~\citep[e.g.][]{akos,sch07a}. 
The $e_{max}$ method uses as an indication of stability a straightforward 
check based on the maximum value of the planet's eccentricity reached during
the total integration time ($T_c$). If the planet's orbit becomes parabolic 
($e_{max} \ge 1$) the system is considered to be unstable.  
The $e_{max}$ is defined as follows:

\begin{equation}
\mathrm{e_{max}} = {\max_{t \le T_{c}}(e(t))}.
\end{equation}

\subsection{ETVs}

Since the first exoplanets in P-Type motion were detected, the investigation
of the eclipse timing variation became more and more important. The ETV signal of 
the secondary star will be induced by an additional planet. This gravitational 
perturbation affects the motions of the two stars and cause their orbits to 
deviate from Keplerian.
In an eclipsing binary, these deviations result in variations over time and 
duration of the eclipse.

This method is particularly important in the case when the planet's orbit 
is not in the line of sight, which causes the absence of a transit signal. 
However, such planets cause perturbations in the orbit of the transiting star,
leading to detectable ETVs. Similar investigations for transit timing variations (TTVs) 
were done in several articles like e.g. \citet{mira05}; \citet{holman05} and \citet{agol07}.
The feasibility of the detection of extrasolar planets by 
the partial occultation on eclipsing binaries was investigated 
by~\citet{schneider90}.
The goal of our work was to show which planet sizes for the S- and P-Type 
configurations are detectable in the ETV signal of the secondary star with 
current observational equipment. In order to approximate the
detectability of possible extrasolar planets by means of ETVs we used the work 
of \citet{syb10} who investigated the sensitivity of the eclipse timing technique 
for the ground and space-based photometric observations.
 They showed in a best-case scenario (excluding e.g. star spots or pulsations), that the 
typical photometric error (detectable timing amplitude dT) for CoRoT is about dT = 4 sec 
for a brightness (L) of 12~[mag] and dT = 16 sec for L=15.5~[mag]. 
Kepler has a dT = 0.5 sec for L=9~[mag] and a dT = 4 sec for L=14.5~[mag].
Future space missions will support the effort to detect smaller planets, like 
for example: 

\begin{itemize}
\item PLATO (Planetary Transits and Oscillations 
of stars) will monitor relatively nearby stars to hunt for Sun-Earth analogue 
systems \citep{rauer14}.
\item TESS (Transiting Exoplanet Survey Satellite) space mission is dedicated to 
detect nearby Earth or super-Earth-size planets on close-in orbits around the 
brightest M dwarfs~\citep{ricker14}. 
\item CHEOPS (Characterising ExOPlanets 
Satellite) will examine transiting exoplanets of known bright and nearby 
host stars~\citep{broeg13}. 
\end{itemize}

For our investigations we will use as detection 
criterion the photometric precision of CoRoT $dT_{crit}=16$~sec 
as well as that of Kepler $dT_{crit}=4$~sec.
We determined the ETVs by calculating the amplitude for the perturbed case, where the 
planet-induced constant rate of apsidal precession is removed by a linear fit. We also 
took into account the long-term effects caused by the binaries motion around the 
systems center of mass and the light travel time effect \citep{mon}.

\section{S-Type}
\label{S}

Several observations of exoplanets in detached binaries motivated us to
investigate the possible detection of exoplanets with ETV's. Therefore we used the 
configuration "primary star-planet-secondary star" with the following initial conditions:

\begin{itemize}
\item {\bf Masses:}
As shown in section~\ref{models} we used different masses for the primary and the secondary star 
(model 1, 2 and 3), which we think represents a quite common mass ratio for binaries and might be 
useful for future observations of different stars (see discussion section~\ref{statistics}).

For the planets we used three different masses: Jupiter ($M_{\jupiter}$), Neptune ($M_{\neptune}$)
and Earth ($M_{\oplus}$).
\item {\bf Semi-major axis:} 
We considered eclipsing binaries, with separations between the stars of $a_{bin}$=0.5, 1 and 3~AU.
The outermost stability border for the possible exoplanets were taken
from the literature \citep{pl02} and verified by numerical integrations. Within the stability
borders we integrated 80 equally distributed configurations in case of $a_{bin}$=0.5~AU and 
$a_{bin}$=1~AU. For $a_{bin}$=3~AU we used 160 equally distributed configurations.

\item {\bf Eccentricity:} The eccentricity of the planet was varied
between 0.0 and 0.5 (and divided into 80 data points). The binary's eccentricity was set 
to 0, 0.2 and 0.4. 
\item All other orbital elements were set to zero ($\omega$~=~$\Omega$~=~M~=~0).
\end{itemize}

\begin{figure*}
\centerline
{\includegraphics[width=9.1cm,angle=0]{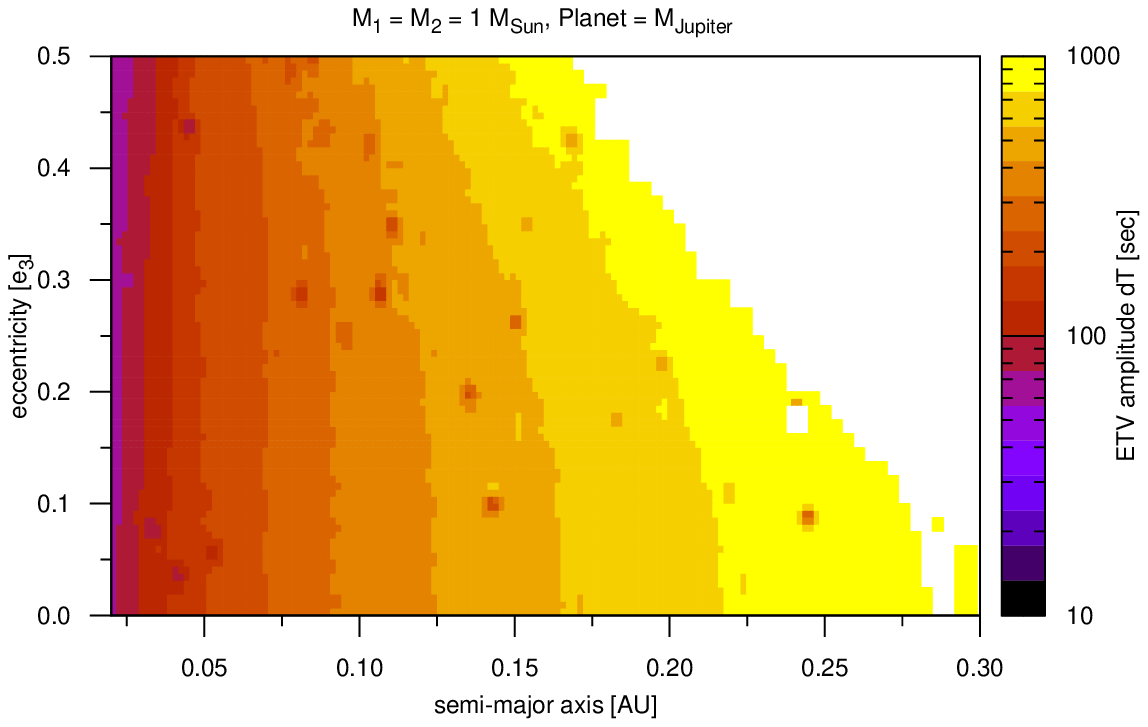}
\includegraphics[width=9.1cm,angle=0]{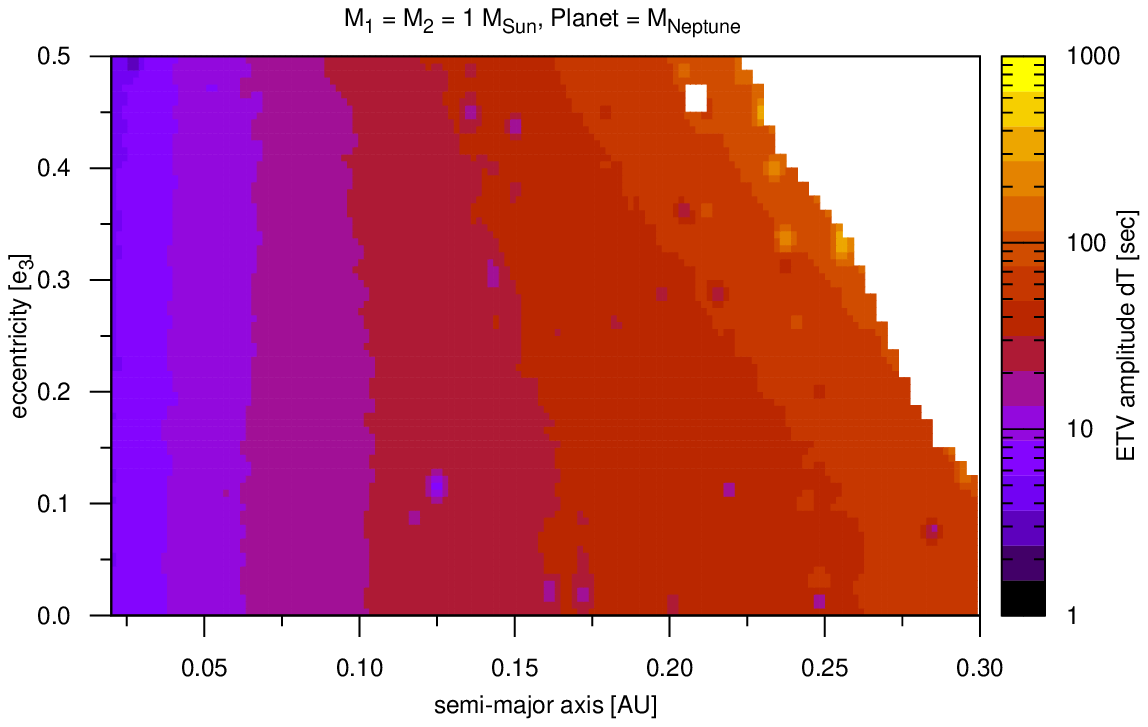}}
{\includegraphics[width=9.1cm,angle=0]{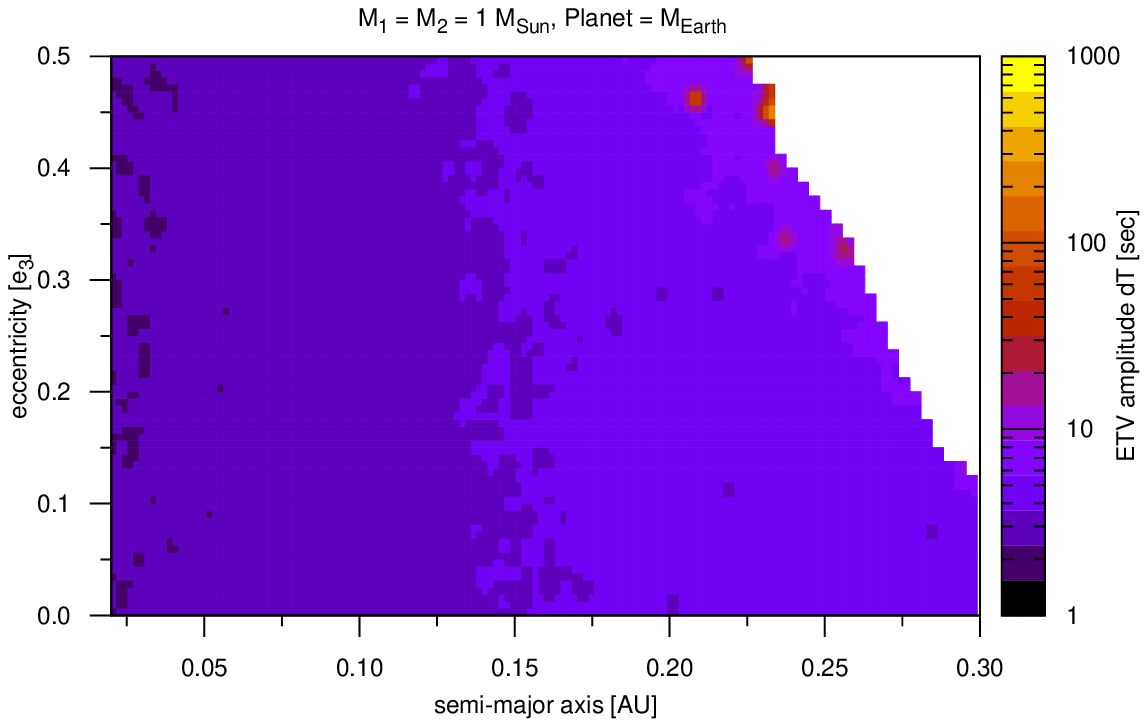}}
\caption{Three different ETV maps in the planar ($i=0^{\circ}$) three-body problem for possible 
exoplanets in S-Type motion for model 1 ($m_1=m_2=1 M_{\odot}$.
For the separation between the two stars we choose $a_{bin}$=1~AU with a mass of the planet of 
1~$M_{\jupiter}$ (upper left graph), 1~$M_{\neptune}$ (upper right graph) and 1~$M_{\earth}$ (lower 
graph). On the x-axes the distance of the planet to the primary star is given and on the y-axes 
its eccentricity. The colour code presents the values of the amplitude of the ETV signal $dT$ in 
[sec]. The yellow and red regions depicts large values of $dT$ whereas the violet and black 
regions represent small ones, whereas the white region corresponds to unstable motion of the 
planet. A colour version of this figure is available in the online version.}
\label{etv1}
\end{figure*}

For our computations of the ETV-maps, we changed the distance from the planet to the primary 
star and the eccentricity of the exoplanet.
The grid size of the ETV maps were changed from 80x80 ($a_3$x$e_3$) for $a_{bin}$ 0.1 and 1~AU and 
extended to 160x80 for $a_{bin}$=3~AU. An example is given in Fig.~\ref{etv1} for model 1 
($M_1=M_2=1$ $M_{\odot}$) for 3 different masses of the planets: 1~$M_{\jupiter}$ (Fig.~\ref{etv1} 
upper left graph), 1~$M_{\neptune}$ (upper right graph) and 1~$M_{\oplus}$ (lower graph). The 
separation of the binaries is $a_{bin}=1$~AU and they have non eccentric orbits. 
The stability border for the separation of $a_{bin}$=1~AU for all planets ($M_{\jupiter}$, 
$M_{\neptune}$ and $M_{\oplus}$) is roughly a=0.3~AU. Close or outside the border, the influence of 
the secondary becomes too large and the planets escape (see Fig.~\ref{etv1} white region). 
As one can see the stability border shrinks 
for higher eccentricities of the planet ($e_3 \ge 0.1$). The timing amplitude dT of planets with 
$M_{\jupiter}$ is 10 times larger ($dt:100-1000s$) than for $M_{\neptune}$ ($dt:10-100s$) and very 
small for the Earth. For the $dT_{crit}=4$~sec it is possible to detect Earth-like planets as well 
as by future space missions, which will have a better time resolution or photometric errors. 
\\
Tables~\ref{stype_nep} and~\ref{stype_earth} summarise our results. The table for planets with 
1~$M_{\jupiter}$ is not shown, because for all stable initial conditions the ETV-maps would be 100 
percent detectable within the typical photometric error of CoRoT ($dT_{crit}=16sec$) and Kepler 
($dT_{crit}=4sec$). 
The Neptune-sized planets are detectable for almost all stable orbits in the ETV-map for 
$dT_{crit}=4sec$, whereas for $dT_{crit}=16sec$ especially the ETVs for small separation of the 
binaries ($a_{bin}$=0.5) and larger eccentricities ($e_{bin}=0.2$ and $e_{bin}=0.4$) are not 
detectable. This is similar for Earth-like planets. However, much more stable orbits are not 
detectable for both photometric errors.    
For $dT_{crit}=16sec$ almost no ETV signals are detectable (only for $a_{bin}=3 AU$ with the model~2
and model~3 for low values of $e_{bin}$).

\begin{table*}
 \center
  \caption{Percent of detectable ETVs for a Neptune-sized planet in S-Type motion.}
  \begin{tabular}{|ccccc|}
  \hline
   $a_{Bin}$ [AU] & $e_{Bin}$ & $m_2$ [$M_{\odot}$] & \% of detectable &  \% of detectable\\
    & & & ETVs [16s] &  ETVs [4s]\\
 \hline
0.5	&	0.01	&	1.0	&	15.04	&	99.90	\\
0.5	&	0.20	&	1.0	&	0.00	&	99.80	\\
0.5	&	0.40	&	1.0	&	0.00	&	95.58	\\
0.5	&	0.01	&	0.5	&	53.67	&	99.95	\\
0.5	&	0.20	&	0.5	&	8.52	&	99.92	\\
0.5	&	0.40	&	0.5	&	0.00	&	99.31	\\
0.5	&	0.01	&	0.1	&	81.84	&	96.41	\\
0.5	&	0.20	&	0.1	&	59.44	&	99.92	\\
0.5	&	0.40	&	0.1	&	29.03	&	99.63	\\
1.0	&	0.01	&	1.0	&	67.78	&	99.86	\\
1.0	&	0.20	&	1.0	&	50.48	&	99.62	\\
1.0	&	0.40	&	1.0	&	6.06	&	99.16	\\
1.0	&	0.01	&	0.5	&	80.96	&	99.53	\\
1.0	&	0.20	&	0.5	&	66.65	&	99.49	\\
1.0	&	0.40	&	0.5	&	33.59	&	99.15	\\
1.0	&	0.01	&	0.1	&	85.35	&	96.08	\\
1.0	&	0.20	&	0.1	&	76.64	&	99.63	\\
1.0	&	0.40	&	0.1	&	52.47	&	97.38	\\
3.0	&	0.01	&	1.0	&	90.04	&	98.87	\\
3.0	&	0.20	&	1.0	&	87.08	&	98.91	\\
3.0	&	0.40	&	1.0	&	78.99	&	98.51	\\
3.0	&	0.01	&	0.5	&	94.61	&	99.09	\\
3.0	&	0.20	&	0.5	&	90.46	&	98.36	\\
3.0	&	0.40	&	0.5	&	83.53	&	98.34	\\
3.0	&	0.01	&	0.1	&	96.49	&	98.68	\\
3.0	&	0.20	&	0.1	&	94.24	&	97.49	\\
3.0	&	0.40	&	0.1	&	88.44	&	97.39	\\
\hline
\end{tabular}
\label{stype_nep}
\end{table*}

\begin{table*}
 \center
  \caption{Percent of detectable ETVs for a Earth-sized planet in S-Type motion.}
  \begin{tabular}{|ccccc|}
  \hline
   $a_{Bin}$ [AU] & $e_{Bin}$ & $m_2$ [$M_{\odot}$] & \% of detectable &  \% of detectable\\
    & & & ETVs [16s] &  ETVs [4s]\\
 \hline
0.5	&	0.01	&	1.0	&	0.00	&	0.00	\\
0.5	&	0.20	&	1.0	&	0.00	&	0.00	\\
0.5	&	0.40	&	1.0	&	0.00	&	0.00	\\
0.5	&	0.01	&	0.5	&	0.00	&	0.00	\\
0.5	&	0.20	&	0.5	&	0.00	&	0.00	\\
0.5	&	0.40	&	0.5	&	0.00	&	0.00	\\
0.5	&	0.01	&	0.1	&	0.00	&	56.00	\\
0.5	&	0.20	&	0.1	&	0.00	&	14.18	\\
0.5	&	0.40	&	0.1	&	0.00	&	0.81	\\
1.0	&	0.01	&	1.0	&	0.00	&	18.27	\\
1.0	&	0.20	&	1.0	&	0.00	&	0.00	\\
1.0	&	0.40	&	1.0	&	0.00	&	0.00	\\
1.0	&	0.01	&	0.5	&	0.00	&	52.15	\\
1.0	&	0.20	&	0.5	&	0.00	&	14.00	\\
1.0	&	0.40	&	0.5	&	0.00	&	0.00	\\
1.0	&	0.01	&	0.1	&	9.95	&	68.44	\\
1.0	&	0.20	&	0.1	&	0.00	&	46.50	\\
1.0	&	0.40	&	0.1	&	0.00	&	19.36	\\
3.0	&	0.01	&	1.0	&	0.00	&	73.32	\\
3.0	&	0.20	&	1.0	&	0.00	&	63.87	\\
3.0	&	0.40	&	1.0	&	0.00	&	39.17	\\
3.0	&	0.01	&	0.5	&	18.82	&	85.84	\\
3.0	&	0.20	&	0.5	&	0.00	&	72.96	\\
3.0	&	0.40	&	0.5	&	0.00	&	53.29	\\
3.0	&	0.01	&	0.1	&	37.30	&	89.08	\\
3.0	&	0.20	&	0.1	&	7.45	&	82.18	\\
3.0	&	0.40	&	0.1	&	0.28	&	66.44	\\
\hline
\end{tabular}
\label{stype_earth}
\end{table*}

\section{P-Type}
\label{P}
For the investigation of the P-Type systems we used the following initial
conditions:

\begin{itemize}
\item {\bf Masses:} The masses of the binary stars were chosen according
to model 1, 2 and 3 (section \ref{models}). The planet in P-Type motion
was integrated with the mass of Jupiter, Neptune and Earth.
\item {\bf Semi-major axis:} For the distances between the two stars we
choose 0.5~AU and 1.0~AU. The innermost distance of the planet needed for
stable motion was taken from literature \citep{schwarz2011} and verified by numerical
integrations. To determine the dependency on the distance we integrated
100 equally distributed configurations up to a distance of 5~AU (for a
distance between the stars of 0.5~AU) respectively 10~AU (for a distance
between the stars of 1.0~AU) from the center of mass of the system.
\item {\bf Eccentricity:} The eccentricity of the planet was varied
between 0.0 and 0.5. The binary's eccentricity was set to 0, 0.2 and 0.4.
\item All other orbital elements were set to zero
($\omega$~=~$\Omega$~=~M~=~0).
\end{itemize}

\begin{figure*}
\centerline{
\includegraphics[width=9.5cm]{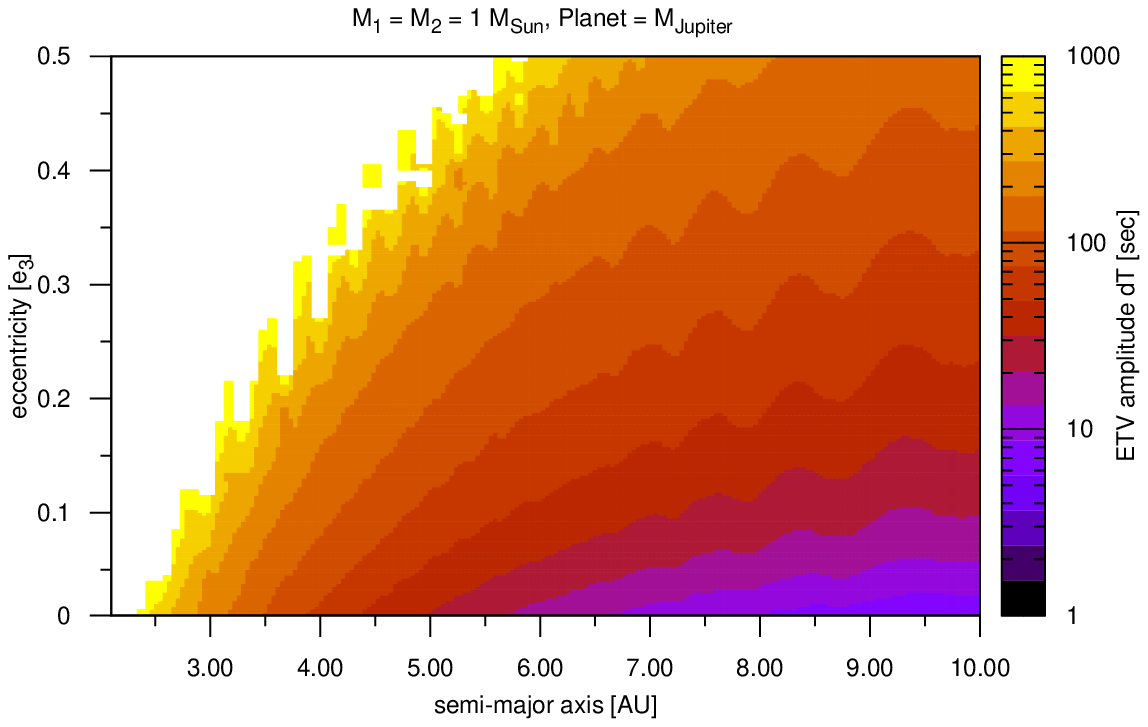}
\includegraphics[width=9.5cm]{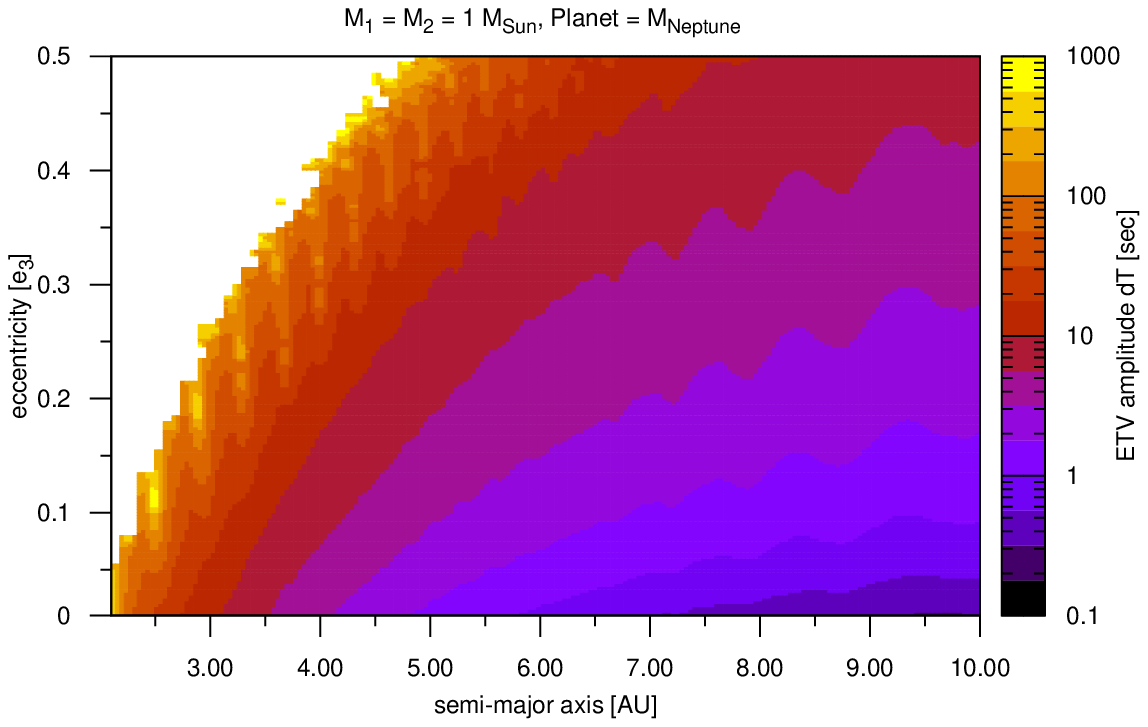}}
\caption{Two different ETV-maps for P-type model~1, a distance between the stars 
of $a_{bin}=1 AU$, the binaries eccentricity of $e_{bin}=0$ and a Jupiter-massed 
planet (left graph) and a Neptune-sized planet (right graph). On the x-axes 
the distance of the planet and on the y-axes its eccentricity is given. 
Resulting in a grid of 100~x~100 initial conditions. The colour code corresponds 
to dT in sec. A colour version of this figure is available in the online version.}
\label{ptypegrid}
\end{figure*}

As an example figure \ref{ptypegrid} shows the results for model~1, a
distance between the stars of 1~AU, a binary eccentricity of 0.0 and a
Jupiter-sized planet (left graph) and Neptune-sized planet (right graph). 
On the x-axes the distance of the planet and on the y-axes its eccentricity 
is given and the axes are subdivided in a grid of 100~x~100 initial
conditions. The colour code corresponds to the ETVs in minutes, where the
white region in the upper left corner corresponds to unstable motion
of the planet.

As one can see Jupiter's ETVs range from approximately 10 to 1000 seconds, which is
much larger than the limits of observability. However, Neptunes ETVs range
from 0.5 to 100 seconds.
Thus we can conclude that Jupiter sized planets could be easily detected in such 
systems, while Neptune sized planets show less strong ETV signals.\\
To give an overview of the results for the different models and initial
conditions we summarise in the following two tables (table \ref{ptype_jup}
and \ref{ptype_nep}) the percentage of detectable ETV Signals for Jupiter-
and Neptune-sized planets. Since different telescopes can reach different
resolutions we give the percentage of all orbits, which show ETV signals
larger than 4~s and larger than 16~s (last two columns in tables~\ref{ptype_jup} 
and \ref{ptype_nep}). In the first three columnes we give the semi-major axis, 
the eccentricity and the mass of the secondary.

\begin{table*}
 \centering
  \caption{Percent of detectable ETVs for a Jupiter-sized planet in P-Type
motion.}
  \begin{tabular}{|ccccc|}
  \hline
   $a_{Bin}$ [AU] & $e_{Bin}$ & M2 [$M_{\odot}$] & \% of detectable &  \% of
detectable\\
    & & & ETVs [16s] &  ETVs [4s]\\
 \hline
   0.5 & 0.01 & 1.0 &  73.44 &  100\\
   0.5 & 0.20 & 1.0 & 100.00 &  100\\
   0.5 & 0.40 & 1.0 & 100.00 &  100\\
   0.5 & 0.01 & 0.5 &  80.29 &  100\\
   0.5 & 0.20 & 0.5 & 100.00 &  100\\
   0.5 & 0.40 & 0.5 & 100.00 &  100\\
   0.5 & 0.01 & 0.1 &  83.95 &  100\\
   0.5 & 0.20 & 0.1 & 100.00 &  100\\
   0.5 & 0.40 & 0.1 & 100.00 &  100\\
   1.0 & 0.01 & 1.0 &  92.00 &  100\\
   1.0 & 0.20 & 1.0 & 100.00 &  100\\
   1.0 & 0.40 & 1.0 & 100.00 &  100\\
   1.0 & 0.01 & 0.5 &  93.14 &  100\\
   1.0 & 0.20 & 0.5 & 100.00 &  100\\
   1.0 & 0.40 & 0.5 & 100.00 &  100\\
   1.0 & 0.01 & 0.1 &  98.31 &  100\\
   1.0 & 0.20 & 0.1 & 100.00 &  100\\
   1.0 & 0.40 & 0.1 & 100.00 &  100\\
\hline
\end{tabular}
\label{ptype_jup}
\end{table*}
\begin{table*}
 \centering
  \caption{Percent of detectable ETVs for a Neptune-sized planet in P-Type
motion.}
  \begin{tabular}{|ccccc|}
  \hline
   $a_{Bin}$ [AU] & $e_{Bin}$ & M2 [$M_{\odot}$] & \% of detectable &  \% of
detectable\\
    & & & ETVs [16s] &  ETVs [4s]\\
 \hline
   0.5 & 0.01 & 1.0 & 0.00  &   0.00\\
   0.5 & 0.20 & 1.0 & 0.00  &  40.59\\
   0.5 & 0.40 & 1.0 & 0.00  &  71.92\\
   0.5 & 0.01 & 0.5 & 0.00  &  21.03\\
   0.5 & 0.20 & 0.5 & 2.24  &  63.67\\
   0.5 & 0.40 & 0.5 & 7.79  &  90.44\\
   0.5 & 0.01 & 0.1 & 0.87  &  31.56\\
   0.5 & 0.20 & 0.1 & 15.00 &  86.43\\
   0.5 & 0.40 & 0.1 & 31.95 & 100.00\\
   1.0 & 0.01 & 1.0 & 0.00  &  47.49\\
   1.0 & 0.20 & 1.0 & 22.36 &  95.54\\
   1.0 & 0.40 & 1.0 & 50.92 & 100.00\\
   1.0 & 0.01 & 0.5 & 1.30  &  59.51\\
   1.0 & 0.20 & 0.5 & 45.35 & 100.00\\
   1.0 & 0.40 & 0.5 & 75.19 & 100.00\\
   1.0 & 0.01 & 0.1 & 13.65 &  70.58\\
   1.0 & 0.20 & 0.1 & 72.02 & 100.00\\
   1.0 & 0.40 & 0.1 & 94.87 & 100.00\\
\hline
\end{tabular}
\label{ptype_nep}
\end{table*}

For Jupiter-sized planets (table \ref{ptype_jup}) all initial conditions
produce ETV signals larger than 4~s and nearly all initial conditions
produce even ETV signals larger than 16~s. For Neptune-sized planets
(table \ref{ptype_nep}) the situation is not that clear, but still nearly
all initial conditions produce a quite large amount of ETV signals above
4~s, while the amount of ETV signals above 16~s shrink clearly for lower
mass planets. The investigation of Earth-sized planets showed almost no
detectable ETV signals. We could find just a few initial conditions (for
$a_{Bin}$ = 1 AU, $e_{Bin}$ = 0.2 or 0.4 and $m_2=0.1 M_{\cdot}$) for which
ETV signals above 4~s could be found and none for ETV signals above
16~s.\\
From our results we can conclude that circumbinary planets down to
Neptune-size can be detected by using ETVs, while even lower massed
planets (i.e. Earth-sized) currently can not be found by ETV signals.

\section{Statistics On Binary Star Systems}
\label{statistics}

The space missions CoRoT and Kepler discovered a huge number of eclipse binaries \citep{maceroni}.
The eclipsing binary frequency of the first CoRoT fields is 1,2\% of all targets. 
A similar, slightly larger, value of 1.4\% is found by \citet{slaw11} and \citet{prsa} for 
the Kepler targets. Most of these binaries have very small periods ($< 10d$), which are not 
interesting for the S-Type configurations.
However, beside these space missions there is a lot of unused data available from former studies, 
collected in several catalogues. Our survey includes also spectroscopic binaries to find 
candidates for our numerical investigations and for the catalogue of exoplanets in binary star 
systems. 

Collecting data and information of binary star systems has by now a long history and 
a considerable amount of catalogues has been published in this context. These databases have 
meanwhile reached a size and complexity, which require not only the consolidation of the existing 
data but also renewed evaluations and statistical analyses.
\par
One of our main challenges is to collect and aggregate the existing data of the semi-major axes 
($a$), the mass ratios ($\mu$), and the eccentricities ($e$) given in former catalogues of binary 
star systems. 
In this process, we concentrate on previous investigations which include publications since 
the 1980s as listed in Tab.~\ref{table:catalogs}.

\begin{table*}
%\centerline
\caption{List of reviewed publications. 
'No.' stands for the total number of stars as offered in the according catalogue. '-' implies 
that no relevant data is available. (\mbox{*}) In these cases, the semi-major axis of the orbit is 
given in arcsec. Due to the lack of data, they could not be converted into AU.}

%\begin{tiny}
\begin{tabular}{lrrrrrrr}
\\
\hline  
 Catalogues                   &  No.  &  $a$ (total)  &  $a$ $<$ 100~AU & $a$ $<$ 20~AU & $a$ $<$ 
3~AU &  $\mu$ or $m$  &  $e$  \\
\hline
%\cite{brancewicz_catalogue_1980}     &  1048  &  x  &  x  &  1048  &  -  \\
%\cite{kraicheva_catalogue_1980}  
%Kraicheva (1980) &  log of $a$      &  -  &  y  &  -  &  -  &     & y\\
\cite{worley_fourth_1983}            &  933 &  932(\mbox{*}) &  -  &  -  &  -  &  933  &  -  \\
\cite{budding_catalogue_1984}        &  414  &  -  &  -  &  -  &  -  &  394  &  -  \\
\cite{corbally_close_1984}           &  170  &  -  &  -  &  -    &  -  &  -  &  -  \\
\cite{pedoussaut_list_1985}          &  1207  &  -  &  -   &  -  &  -  &  -  &  -  \\
\cite{pedoussaut_spectroscopic_1988} &  436  &  310  &  254 &  170  &  75  & 253  & 421  \\
\cite{malkov_catalogue_1993}         &  288  &  -  &  -  &  -  &  -  &  287  &  -  \\
\cite{perevozkina_catalog_1999}      &  44  &  44  &  44  &  44  &  44  &  -  &  -  \\
\cite{svechnikov_catalog_1999}       &  113  &  113  &  113  &  113  &  111  &  -  &  -  \\
\cite{liu_catalogue_2001}            &  280  &  -  &  -  &  -  &  -  &  -  &  -  \\
\cite{downes_catalog_2001}           &  1314  &  -  &  -  &  -  &  -  &  -  & -  \\
\cite{mason01}                       &  132120  &  78508(\mbox{*})  &  -  &  -  &  -  &  -  &  -  \\
\cite{udalski_optical_2002}          &  177  &  -  &  -  &  -  &  -  &  -  &  -  \\
\cite{pribulla_catalogue_2003}       &  361  &  -  &  -  &  -  &  -  &  116  &  -  \\
\cite{surkova_vizier_2004}           &  232  &  323  &  232  &  232  &  230  &  -  &  - \\
\cite{pourbaix_sb9_2004}             &  4031  &  -  &  -  &  -  &  -  &  -  & 4031  \\
\cite{mermilliod_red_2007}           &  157  &  155  &  96  &  24  &  2  &  -  &  157  \\
\cite{ritter_catalogue_2003,ritterh._vizier_2011}          &  2072  &  -  &  -  &  -  &  -  &  281  &  -  \\
\cite{mace_vizier_2014}              &  1565  &  -  &  -  &  -  &  -  &  -  &  -  \\
\cite{nicholson_vizier_2015}         &  9450  &  -  &  -  &  -  &  -  &  -  &  -  \\
\hline
\end{tabular}
%\end{tiny}
\label{table:catalogs}
\end{table*}

An early example of a comprehensive data collection on binary star systems includes the work of 
\cite{worley_fourth_1983} which is based on the ``Finsen-Worley Catalogue" published in 
\citeyear{finsen_finsen_1970}. The ``4th Catalogue of Orbits of Visual Binaries" contains orbital 
elements of about 930 objects in 847 systems, whereby triples are counted as two systems.  
The statistical distribution of the semi-major axis illustrates that almost 96\% of the visual 
binaries are located at a angular separation of less than 5~arcsec.
\par
In 1988 and 1989, the ``15th Complementary Catalogue of SBs" was published by 
\citeauthor{pedoussaut_spectroscopic_1988} and \citeauthor{malkov_errata_1989}, respectively. 
This database contains the orbital data and the derived masses of 436 spectroscopic binaries. 
The statistical analysis of the available data of 310 semi-major axes shows that 
just above 92\% of the stars have an $a\cdot sin(i)$ $<$ 200~AU (minimum distance).

\par

The ``Catalogue of eclipsing binaries parameters" of \cite{perevozkina_catalog_1999} and 
\cite{perevozkina_vizier_2004} respectively, not only includes orbital parameters, masses, and 
luminosities but also photometric orbit data of 44 eclipsing binary systems whereby all values of 
$a$ are smaller than 0.5~AU.
\par
\begin{figure}
      \centerline{
      \includegraphics*[width=6cm,angle=270]{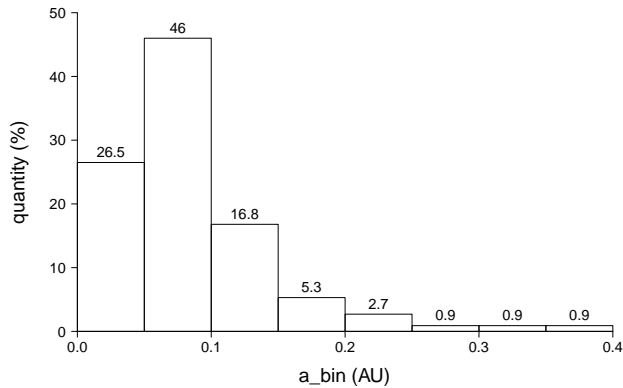}}
      \caption{Semi-major axis of 113 DMS-type eclipsing binaries 
\protect\citep{svechnikov_catalog_1999}}
    \label{fig:a_bin_1999_Svechnikov}
\end{figure}

\cite{svechnikov_catalog_1999} and \cite{svechnikov_vizier_2004} respectively, published the 
"Catalogue of DMS-type eclipsing binaries" which contains information of 113 binaries with 
photometric and spectroscopic parameters. The semi-major axis of the detached main-sequence-type 
eclipsing binaries is illustrated in Fig.~\ref{fig:a_bin_1999_Svechnikov}.

\par

The catalogue of ``Semi-detached eclipsing binaries" of \citeauthor{surkova_vizier_2004} from 
the year \citeyear{surkova_vizier_2004} is a collection of slightly more than 230 semi-detached 
eclipsing binary star systems with known photometrical orbital elements. The distribution of the 
semi-major axes illustrates a significant accumulation of systems with an $a$ smaller than 0.1~AU. 
Just over 85\% of the eclipsing binaries are located in that area.

\begin{figure}
      \centerline{
      \includegraphics*[width=6cm,angle=270]{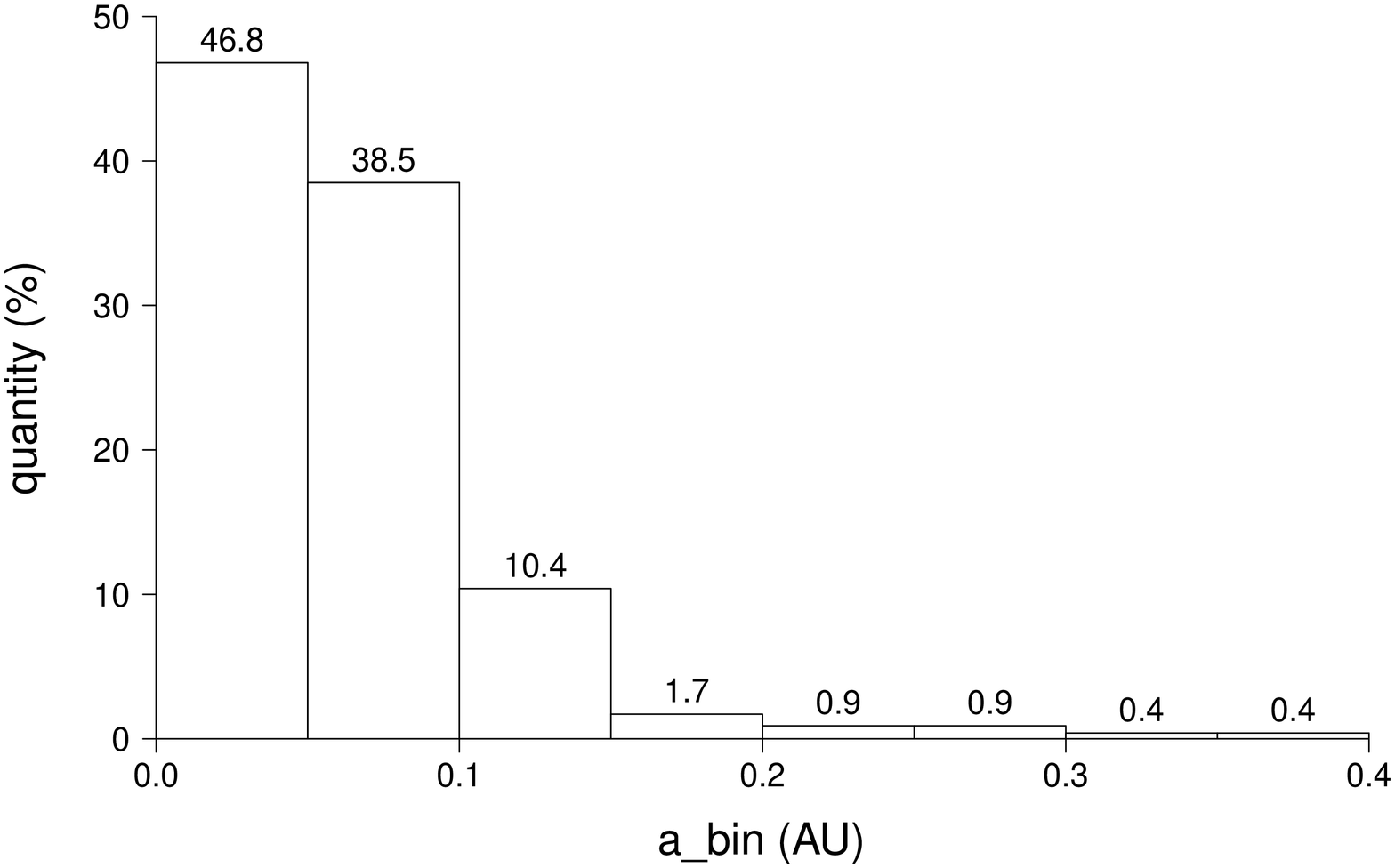}}
      \caption{Semi-major axis of 231 eclipsing binaries \protect\cite{surkova_vizier_2004}}
    \label{fig:a_bin_2004_Surkova}
\end{figure}

\begin{figure*}
\centerline
{\includegraphics*[width=6.5cm,angle=270]{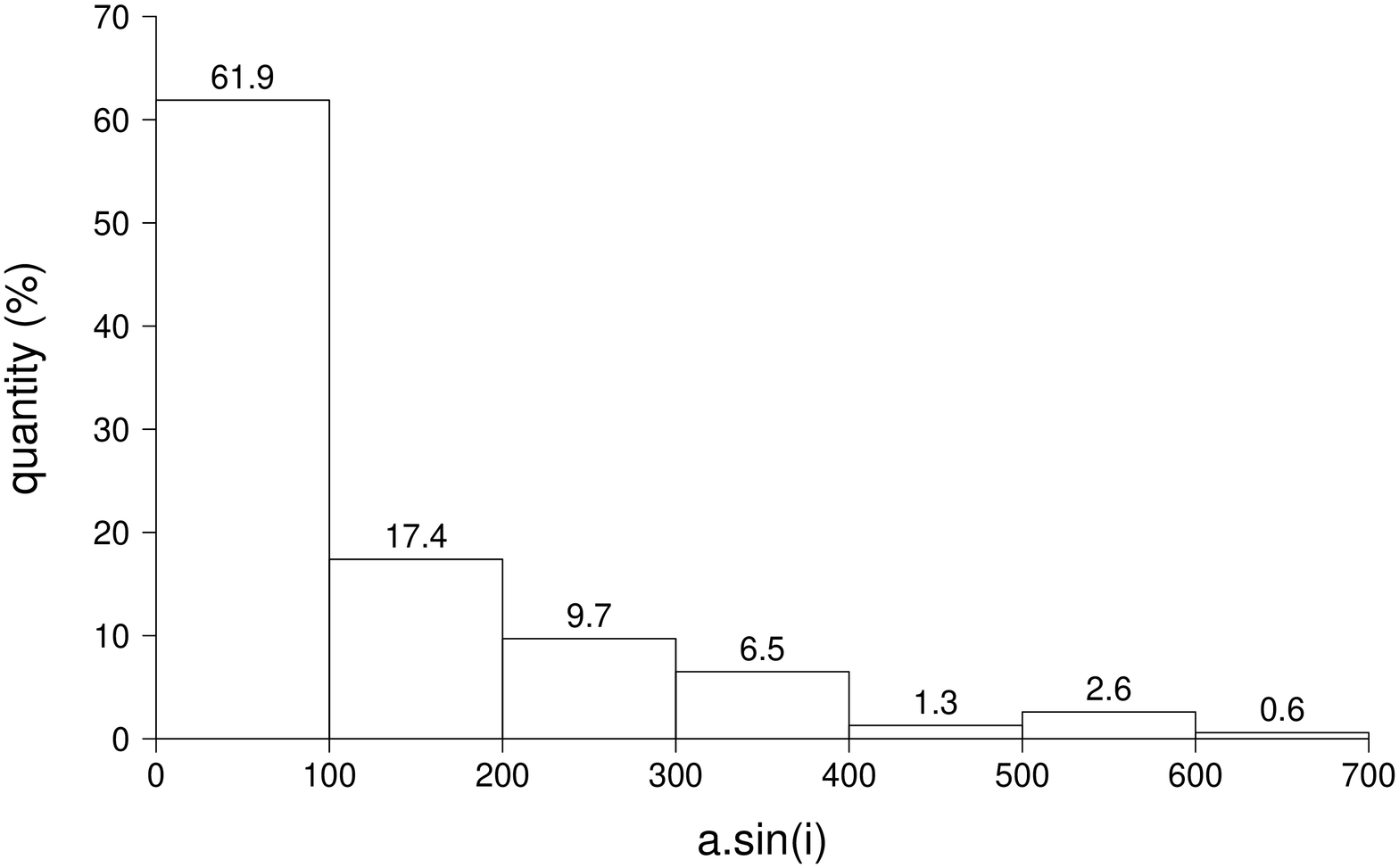}
\includegraphics*[width=6.5cm,angle=270]{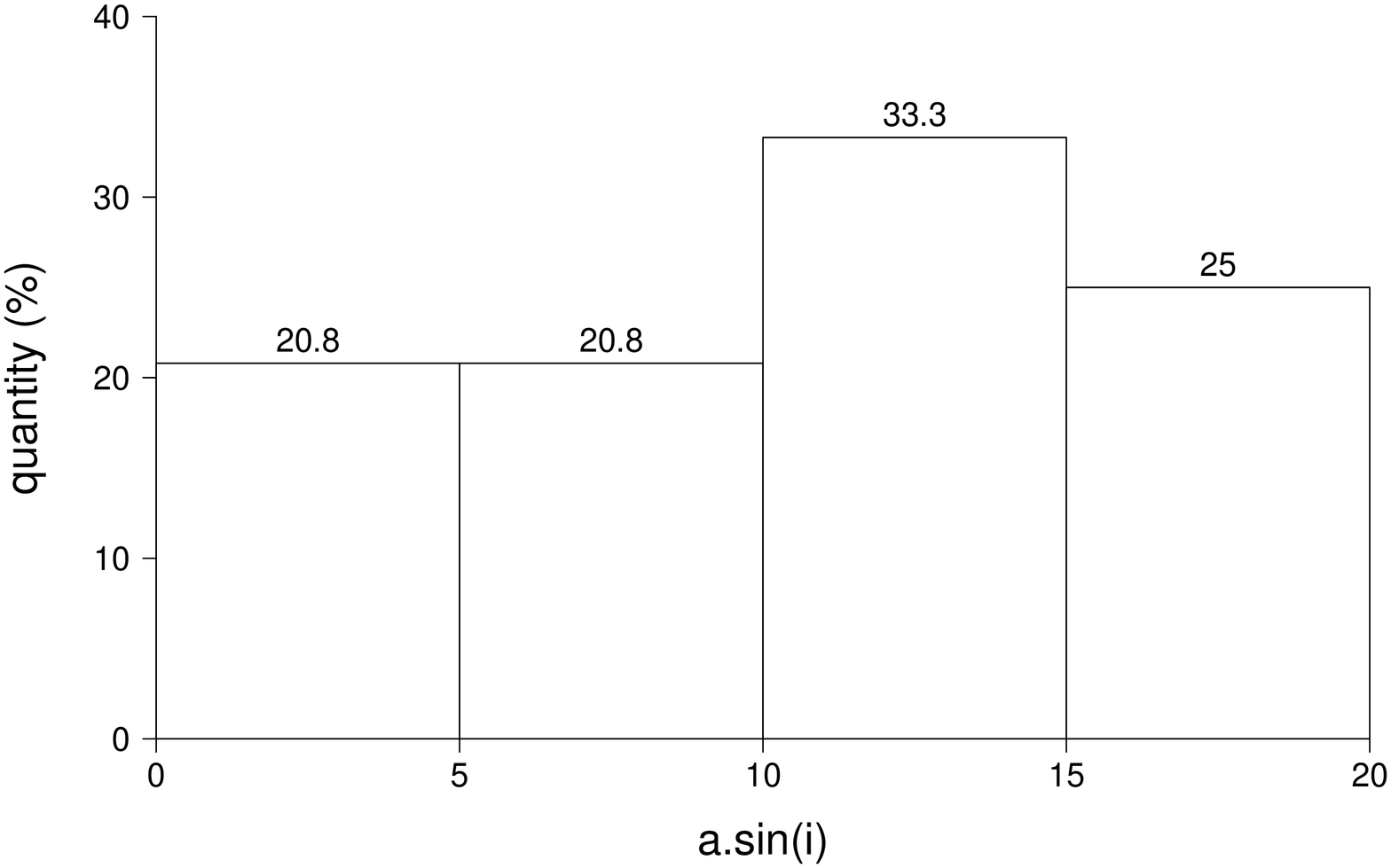}}
\caption{Left-hand panel: Semi-major axis of 155 spectroscopic binaries 
\protect\citep{mermilliod_red_2007}. Right-hand panel: Semi-major axis of 24 spectroscopic 
binaries with $a$ $<$ 20~AU 
\protect\citep{mermilliod_red_2007}.}
\label{fig:2007_Mermilliod}
\end{figure*}

The semi-major axis of 155 spectroscopic binaries with red-giant primaries in open clusters 
\citep{mermilliod_red_2007} are shown in the left-hand panel of Fig.~\ref{fig:2007_Mermilliod}. 
Orbital periods range from 2.07 to 689 days (1.89 years). 
It is apparent from this distribution that almost 62\% of the spectroscopic binaries have a 
semi-major axis smaller than 100~AU with a nearly exponential increase of binaries towards smaller 
$a$. The statistical evaluation for $a$ $<$ 20~AU is presented in the right-hand panel of 
Fig.~\ref{fig:2007_Mermilliod}. We found out that most of the detected exoplanets in binary-star 
systems are stars with masses like our sun or slightly smaller (as shown in the catalogue of 
exoplanets in binary star systems). But, \citet{marsh11} showed that close pairs of white dwarfs 
are very common in our Galaxy, with the order of 100-300 million.

\par

It is of particular interest for our study to find binaries within a separation of $a$ $<$ 3~AU. 
Based on our statistics, we have found a set of 462 candidates that fit our requirements.
With regard to the planning of possible future catalogues, a uniform classification of the 
semi-major axis (e.g. in AU) would be preferable in order to evaluate statistical analyses and to 
provide useful information.
\par

\section{CATALOGUE OF EXOPLANETS IN BINARY STAR SYSTEMS}
\subsection{Motivation for creating the catalogue}

The Extrasolar Planets catalogue was the first online catalogue and is available
since February 1995 at http://exoplanet.eu \citep{mart04}.
The catalogue has been upgraded in 2005 by additional graphical and statistical online
services \citep{le07}.
Other databases followed some years later: the California and Carnegie
Planet Search table at http://exoplanet.org \citep{but06} and the Geneva 
Extrasolar Planet Search Programmes table (Mayor, Queloz, Udry and Naef) providing
first hand data from the observers using the radial velocity and transit method.
The advantage of the Extrasolar Planets Encyclopaedia \citep{schneider11} -- 
maintained by the exoplanet TEAM -- is that it lists all detection methods (astrometry,
pulsar timing, microlensing, imaging etc.).

Cataloguing the data of exoplanetary systems becomes more and more important, due to the 
fact that they conclude the observations and support the theoretical studies.

Since planets in binary star systems were detected they become more important. 
In 2013 we started to compile a catalogue for binary and multiple star systems because at that 
time there did not exist a list of exoplanets in binary star systems.
Now also the Open Exoplanet Catalogue shows exoplanets in binary and
multiple star systems, which is a community driven and decentralised astronomical 
database and available at http://www.openexoplanetcatalogue.com/ \citep{rein12}.
At the beginning of our catalogue we wanted to supplement the ``Extrasolar Planets 
Encyclopedia'', in agreement and with the support of J. Schneider and his team.
In case of binary and multiple star systems the challenge is big due to
the fact that the observations are more complicated and the data have much more errors than
for single star systems.
What concerns us primarily are the statistics, that is why we do not present the errors in 
our list. If more details are needed, we made a link to the Extrasolar Planets
Encyclopedia (the link is contained in the name of the system).
Another purpose of our catalogue is to present review statistics of other
binary catalogues, which is a big challenge because the catalogues present only very special 
stars or regions of our galaxy and are non-uniform.

\subsection{Binary star systems}
The catalogue is described as it was in the year 2016, organised in 12 columns and will be 
updated monthly.
We distinguish detection from discovery, because some planets for example are discovered by
radial velocity and detected by transit afterwards.

One can sort in two directions: ascending, meaning from the lowest value to the highest, or 
descending. 
For example, by clicking on the header e.g. discovery the list will
be sorted after the largest value, when you click again it will be sorted after the smallest 
value. The list is originally sorted by the distance between the binaries ($a_{bin}$), all
rows can be sorted in the same way except the comments. 
In addition an introduction and help is also given in the menu bar of the catalogue including an 
example list. 
To make your own statistics the data is available as .csv file.
All systems are linked to The Extrasolar Planets Encyclopedia some of the systems to the 
Open Exoplanet Catalogue where one can find references and additional data on the systems. 

\subsubsection{Star Data}
\begin{figure*}
\centerline
{\includegraphics[width=16.8cm,angle=0]{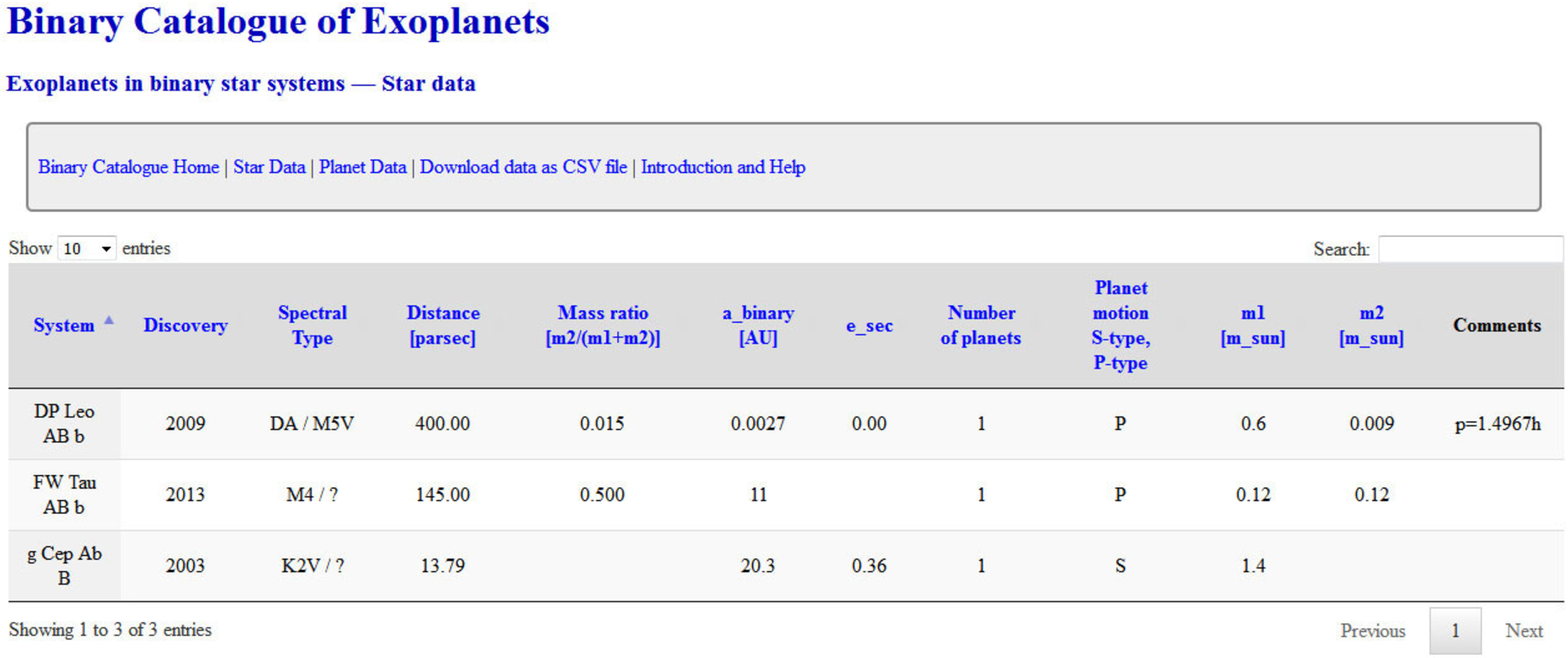}}
\caption{Excerpt from the table of discovered systems with the known data for the stars.
A colour version of this figure is available in the online version. }
\label{star}
\end{figure*}

This part of the catalogue represents only the stellar data of the system (see Fig.~\ref{star}).\\
{\bf System}\\
Name or designation of the system and the structure of the system, where capital letters refer to 
a star, and small letters refer to a planet. Example: DP Leo AB b
"Ab B" or "A Bb" referred to a S-Type planet, while "AB b" refer to a P-Type planet as marked in 
the column on the planetary motion. \\
{\bf Discovery}\\
Gives the year of the first discovery. \\
{\bf Spectral type}\\
This shows the spectral types of the stars. Unfortunately the data for some systems is incomplete. 
\\
{\bf Distance [parsec]}\\
Distance from the Sun to the system in units of parsecs (1 parsec = 3.26 light-years).\\
{\bf Mass ratio ($\mu$})\\
Given as dimensionless proportion $\mu ={\frac {m_2} {(m_1+m_2)}}$, where $m_1$ is the mass of the 
first star and $m_2$ is the secondary star's mass. \\
{\bf a$_{binary}$ [AU]}\\
Represents the distance between the double stars given in astronomical units.
If the semi-major is not given, a minimum of a$_{binary}$ will be approximated trigonometrically by
 the published separation angle $\alpha$ given in [arcsec] and the distance from the Sun to system 
$d$ given in [parsec]. $a={d} \cdot tan{\alpha}$
\\
{\bf Eccentricity (e$_{sec}$})\\
Represents the eccentricity of the second star. This parameter is very rarely known.\\
{\bf Number of planets}\\
Systems with one planet are dominant, but multiplanet systems become more and more frequent.\\
{\bf Planet motion S-type, P-type}\\
As shown in the introduction see also Fig.~\ref{fig1}.\\
{\bf Mass: $m_1$ [$M_{\odot}$] and $m_2$ [$M_{\odot}$]}\\
Mass of the first and the second star given in units of the masses of our Sun.\\

\subsubsection{Planet Data}
\begin{figure*}
\centerline
{\includegraphics[width=16.8cm,angle=0]{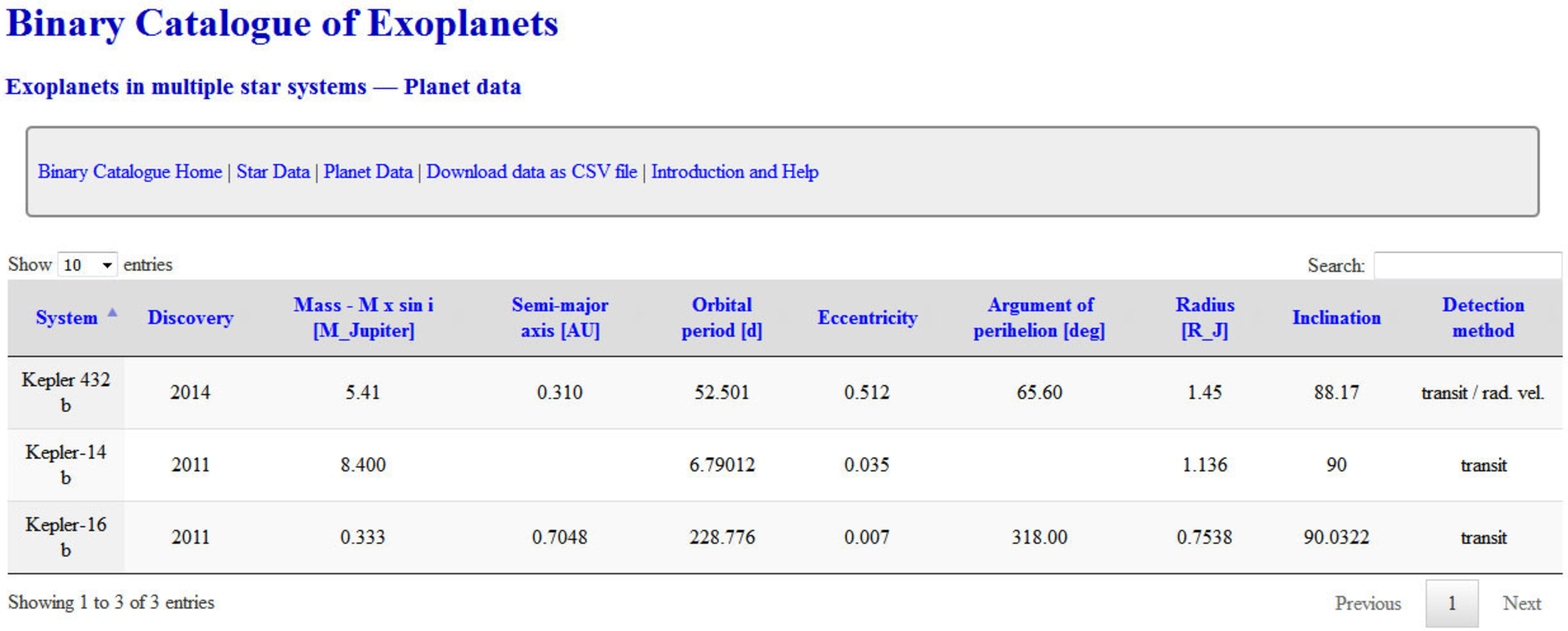}}
\caption{Excerpt from the table of discovered systems with the known data for the planets.
A colour version of this figure is available in the online version.}
\label{planet}
\end{figure*}

Here we present the data of the planets, where the first two columns are similar to the star data 
(see Fig.~\ref{planet}).\\
{\bf Mass M x sin i}\\
The portion of a distant planet's mass that is detectable is determined by its line of sight, when 
observed from Earth. If the angle of inclination from the "face-on" position is "i", then the 
component which is in line with the Earth is given by sin(i).\\
{\bf Semi-major axis [AU]}\\
Represents the semi-major axis of the planet's orbit given in astronomical units. If the semi-major
is not given, it will be derived from the published orbital period and from the mass of the host 
star through the Kepler law.  $a=\sqrt[3] {\frac {G \cdot M_{*} \cdot P^2}  {4 \cdot \pi^{2}}}$  \\
For S-type: $M_{*}$ is the mass $m_1$ of star~1 or $m_2$ of star~2 depending on the planet's orbit.
For P-type: $M_{*}$ is the sum of the masses $m_1$ and $m_2$. These approximations are strongly 
influenced by the star's masses. \\
{\bf Orbital period [d]}. \\
Represents the orbital period of the planet given in days. \\
{\bf Eccentricity}\\
Represents the eccentricity of the planet. \\
{\bf Argument of perihelion [deg]}\\
Represents the angle from the body's ascending node to its periapsis, measured in the direction of 
motion.\\ 
{\bf Radius [$R_J$]}\\
Represents the planet's radius given in units of one Jupiter radius.\\
{\bf Inclination}\\
This value does not always represent the orbital inclination of the planet, especially for 
transiting planets it shows only the inclination relative to the line of sight.\\
{\bf Detection method}\\
Shows the different detection methods which were used for the observations.\\

\subsection{Multiple star systems}
\label{multi}
\begin{figure*}
\centerline
{\includegraphics[width=12.1cm,angle=0]{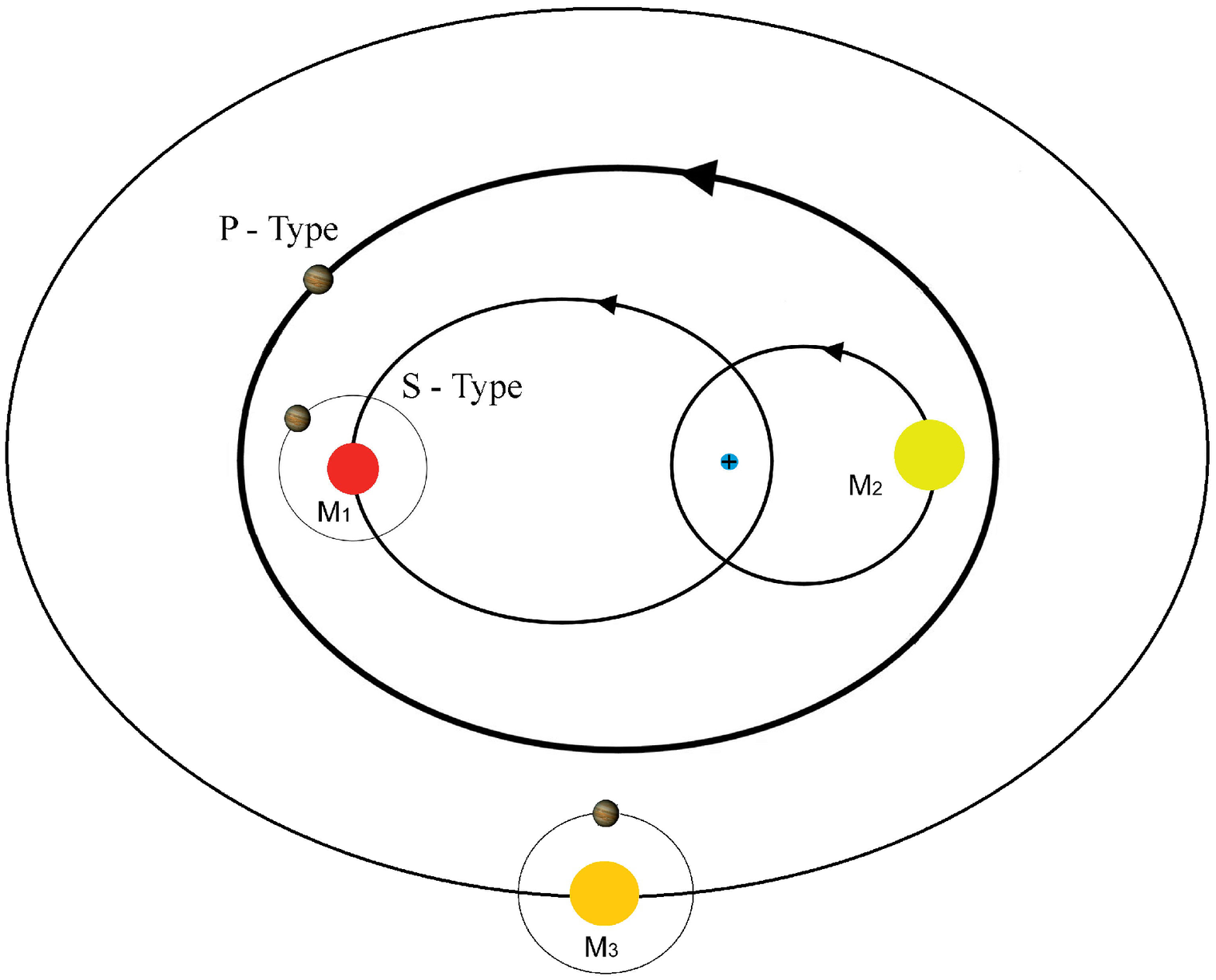}}
\caption{Scheme of the different dynamical possibilities of exoplanets in triple star systems. 
The symbol "+" marks the center of mass of the system. A colour version of this figure is 
available in the online version.}
\label{triple}
\end{figure*}

\begin{figure*}
\centerline
{\includegraphics[width=16.1cm,angle=0]{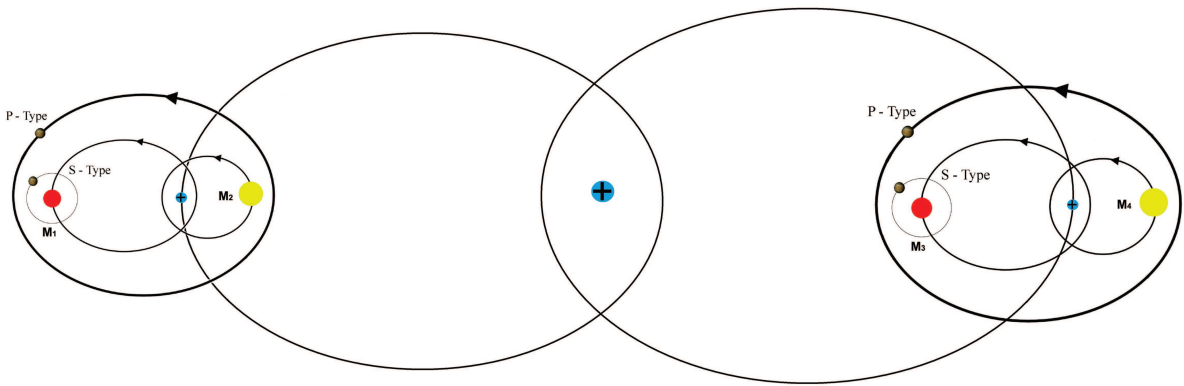}}
\caption{Scheme of the different dynamical possibilities of exoplanets in quadruple star systems. 
The symbol "+" marks the center of mass of the system. A colour version of this figure is 
available in the online version.}
\label{quad}
\end{figure*}

Beside binary star systems also multiple star systems may harbour exoplanets. The different 
possibilities for triple star systems are shown in Fig.\ref{triple}, whereas quadruple star 
systems are presented in Fig.\ref{quad}.

\subsubsection{Star Data}
In this list the first three columns are similar to the list of the binary star systems.\\
{\bf a$_{triple}$ or a$_{quadruple}$ [AU]}\\
Distance in AU of the third star from the inner binary, or of the two binaries from each other.\\
{\bf a$_{binary1}$ [AU]}\\
Separation of the inner binary in case of a triple star system.\\
{\bf a$_{binary2}$ [AU]}\\
Separation of the other binary in case of a quadruple star system.\\
{\bf Number of planets}\\
Systems with one planet are dominant, but multiplanet systems become more and more frequent.  \\
{\bf Number of stars}\\
Total number of detected stars in a multiple star system.\\
{\bf Mass of $m_1$ - $m_4$ [$M_{\odot}$]}\\
Mass of the $1^{st}$, $2^{nd}$, $3^{rd}$ and $4^{th}$ star given in units of the mass of our Sun. 

\subsubsection{Planet Data}
The list is completely identical with the planet data list of binary star systems.

\section{Conclusions}
In this article we used the statistics of the binary catalogue of exoplanets, which we introduce 
in the appendix. 
We prepared statistics of exoplanets in well detached binary systems. In the second part of the 
article we enlarged the statistics by the investigation of well detached binary star systems 
from several catalogues and discussed the possibility of further candidates.
Finally we investigated the possibility to detect exoplanets in well detached binary systems
with eclipse timing variations.

In the statistics of the binary catalogue of exoplanets we could show that the separation
which we used for our calculations are not only of theoretical interest (Fig.~\ref{fig2}, 
lower graph). This also applies for the mass ratios which we used in the models 1,2 and 3 
(see section~\ref{models} and Fig.~\ref{fig2}, upper graph).
We enlarged our investigation with further studies of well detached binary star systems from 
several catalogues and discussed the possibility of further candidates.
These investigations resulted in 462 candidates have star separations not larger 
than 3~AU. This is the separation which we investigated in the ETV study.

In this paper we studied the circumstances favourable to detect S- and P-Type 
planets in well detached binary-star-systems using eclipse timing variations (ETVs). 
To determine the probability of the detection of such variations with ground 
based telescopes and space telescopes, we investigated the dynamics of well detached 
binary star systems with a star separation in the range of $0.5 \le a_{bin} \le 3$AU.
We performed numerical simulations by using the full three-body problem 
as dynamical model. The stability and the ETVs are investigated by computing 
ETV maps for different masses of the secondary star (model 1-3), separations 
($a_{bin}=0.5,\,1\,$ and $3$~AU) and eccentricities ($e_2$=0, 0.2 and 0.4). In addition 
we changed the planet's mass (Earth, Neptune and Jupiter size) eccentricities ($e_3$=0-0.5) 
and semi-major axis (depending on the configuration S- or P-type).
For our investigations we used as detection criterion the photometric precision 
of CoRoT $dT_{crit}=16$ sec as well as that of Kepler $dT_{crit}=4$ sec, which we think
is a realistic limit. 
In general the ETV amplitude $dT$ depends mainly on the eccentricity, the semi-major axis
and the mass of the planet. The stars separation and eccentricity ($e_2$) mainly restricts the 
stable region of the planets.
We conclude that many amplitudes of ETVs are large enough to detect exoplanets -- 
with Neptune and Jupiter-sizes -- in S-type and P-type configurations. 
Whereas for the S-type configuration also Earth-size planets provide detectable ETV signals. 
\par
We can conclude that possible terrestrial-like planets are detectable in binary star systems 
by the help of eclipse timing variations with restrictions. However, future space missions 
will have a better precision which will enlarge the number of detectable ETV signals.

\section*{Acknowledgments}
R. Schwarz, B. Funk and \'A. Bazs\'o want to acknowledge the support by the Austrian FWF project 
P23810-N16.

%\appendix

%\section[]{Large gaps}

%(This appendix was not part of the original paper by
%A.V.~Raveendran and is included here just for illustrative
%purposes. The references are not relevant to the text of the
%appendix, they are references from the bibliography used to
%illustrate text before and after citations.)

\end{document}